# Earth-Projected Clustering of Historical Optical Transients in the Palomar Observatory Sky Survey-I (POSS-I)

Stephen Bruehl[a], Beatriz Villarroel[b], Hichem Guergouri[c], Brian Doherty[d] and Alina Streblyanska[e]

[a]Department of Anesthesiology, Vanderbilt University Medical Center, 701 Medical Arts Building, 1211 Twenty-First Avenue South, Nashville, TN 37212, USA

[b]Nordita, KTH Royal Institute of Technology and Stockholm University, Hannes Alfvéns väg 12, 106 91, Stockholm, Sweden

[c]Research Unit in Scientific Culture and Mediation- CERIST Constantine, 25016, Algeria

[d]Independent Researcher, Dallas, TX, USA

[e]Instituto de Astrofísica de Canarias, Avda Vía Láctea S/N, La Laguna, E-38205, Tenerife, Spain

CORRESPONDING AUTHOR: Beatriz Villarroel, Ph.D, Nordita, KTH Royal Institute of Technology and Stockholm University, Hannes Alfvéns väg 12, 106 91, Stockholm, Sweden

E-Mail: beatriz.villarroel@su.se

Abstract

Palomar Observatory Sky Survey-I (POSS-I) images obtained before the launch of Sputnik contain star-like point sources consistent with sub-second flashes that are absent from all subsequent observations. The nature and origin of these transients remain poorly understood. Given models suggesting they are in geosynchronous orbit, we estimated the latitude/longitude over which POSS-I transients appeared. From a dataset of 107,875 POSS-I transients, we selected a subset with high probability of representing real objects based on a machine learning model. Two independent approaches were used to test whether transients non-randomly grouped over specific locations (hotspots). Based on comparing transient numbers in discrete 5° × 5° Earth-projected regions to numbers randomly expected in these same regions, 16 statistically-significant hotspots were identified. Separately, two-step cluster analysis identified 6 highly-distinct hotspots (silhouette value = 0.70). The two methods identified several similar hotspots not attributable to POSS-I survey biases: 1) Pacific Ocean west of southern Mexico/Central America, 2) southern Gulf of Mexico, 3) southwestern US (Sedona, AZ, White Sands, NM regions). Transients accounting for reported transient-nuclear testing associations displayed location specificity ($p<.0001$), with exclusively Pacific transient locations during Pacific nuclear testing and primarily southwestern US locations during Nevada testing. Findings hint at intriguing POSS-I transient characteristics.

## Introduction

Transient, star-like phenomena have been identified in sequential images over short timescales in the Palomar Observatory Sky Survey (POSS-I) and other historical sky surveys [1-6]. These transients are absent in images taken immediately before they appear and in all images from later surveys[1-4]; they do not correspond with known objects in optical catalogs (e.g. Gaia, Pan-STARRS). They display compact, circular point-spread functions and morphology consistent with brief flashes (specular reflections) lasting from sub-second to a few seconds[7,8]. Such flashes cannot be due to artificial satellites as all images were obtained before the launch of Sputnik in October of 1957.

We have previously reported 107,875 POSS-I transients identified via an automated pipeline[1,9]. Because these transients were observed over 70 years ago, understanding their nature and origin presents challenges. Replication in modern sky surveys is made difficult by the large number of satellites and orbital debris producing similar flashes. Distinguishing astronomical transients from plate defects (hairs, dust, emulsion defects) on scanned archival photographic plates is an additional barrier. Despite these challenges, several complementary approaches have recently been used to better characterize transient characteristics[10-12].

To address issues with false positives due to plate defects, we developed a machine learning (ML) model using a supervised learning approach [10]. The model accurately distinguished between plate defects and astronomical transients (area under the curve = 0.81) and it assigned a probability value to each POSS-I transient indicating likelihood that it represented a real astronomical object rather than plate defects[10]. Intriguingly, previous findings that transients are more likely to appear within +/- 1 day of historical above ground nuclear tests[13] were confirmed in this ML study, with associations strongest in transients assigned the highest probability of being real objects (probability $\geq 0.90$) [10]. Moreover, the previously reported deficit in transients in the Earth's shadow (shadow deficit) [14] was also found to be largest in the highest probability transients [10]. Both findings support high probability transients as being real astronomical phenomena.

Because the shadow deficit is consistent with transients representing reflective objects in Earth orbit and recent findings suggest transients may be due to brief flashes like tumbling objects in orbit exhibit [12.14], we began exploring their potential orbital parameters. We recently developed a modelling approach that permits estimating the altitude of POSS-I transients based on shadow deficit characteristics [12]. After filtering for high probability transients, results suggested orbital altitudes between 20,000 km - 35,000 km above Earth, with the largest shadow deficit at approximately geosynchronous altitude [12]. The latter estimate was independently supported by findings of a significant excess of transients at low declinations [12], consistent with observations of modern geosynchronous satellites [15].

Building on these recent findings, the current work explored the Earth-projected locations over which transients were observed. Specifically, given an assumed geosynchronous orbital altitude (≈35,786 km), combined with information regarding the Mount Palomar observatory location and observation parameters (e.g., right ascension, declination), we estimated the latitude and longitude over which transients were positioned at the time POSS-I transient images were taken.

We then applied two independent statistical approaches to determine whether the observed transient locations represented a random distribution, or rather, distinct clustering over specific locations on Earth (for convenience, hereafter referred to as “hotspots”). In the first approach, we identified 5° × 5° location bins with an excess of high probability transients relative to a random location control sample. In the second approach, we used cluster analysis, a statistical pattern recognition technique, to test for non-random clustering of locations of high probability transients. Finally, we compared and contrasted findings based on the two methods. Hotspots appearing across both analytic methods and not in random location data provided convergent validity for the identified hotspots.

## Results

### Transient Excess Analysis Results

We first tested for locations showing a significant excess of transients relative to plate-aware control data reflecting random locations with the same time and sky coverage as the POSS-I survey. We binned both the transient and control samples into identical 5° × 5° cells projected onto the Earth, each representing different geographical regions. We then compared the observed number of transients in each cell to the number randomly expected based on a uniform distribution on each plate used in the survey, controlling for differences in survey coverage across plates (see Methods for details). The transient sample was restricted to transients with probability ≥ 0.70 of being a real object according to our ML model [10].

Of 137 control cells (5° × 5° each) within which at least 5 transients were expected, 16 cells (11.7%) exhibited a significant excess of transients relative to random control expectations based on bonferroni-corrected conditional binomial tests (Table 1). These 16 hotspots are presented in geographical context in Figure 1. Hotspot 15 is centered at 107.5° W, 32.5° N, 35 km northeast of Deming, NM, 125 km west of White Sands, NM, and 125 km northwest of El Paso, TX; all three locations fall within the same 5° × 5° cell. Hotspots 4, 7 and 8 lie over or near Tampico-Vera Cruz, Oaxaca and the Tabasco–Chiapas region of southern Mexico, while hotspot 5 lies offshore in the Bay of Campeche. Together, these hotspots form a concentration extending from southern Mexico into the southern Gulf of Mexico. Three hotspots showing the greatest transient excess (Hotspots 1-3) formed a concentration in the eastern Pacific, off the west coast of southern Mexico and Central America. Hotspot 9 lies approximately at the latitude of Baja California Sur. Although the control sample itself is biased towards these Pacific regions owing to greater survey coverage in this area, excess analysis shows these transient concentrations cannot be explained by this bias alone; a significant excess remains relative to plate-aware controls even after Bonferroni adjustment for multiple tests.

To determine whether hotspots were driven by individual photographic plates (e.g. local clustering of plate defects), we decomposed each significant cell by plate and calculated transient excess separately for every contributing plate (see Methods). Fourteen of the 16 significant 5° × 5° cells showed a transient excess beyond expectations on at least two different plates, and 12 showed excess on at least three plates. Only the two adjacent cells centered at 92.5° W (Hotspots 5 & 8) were associated exclusively with a single plate, XE435. When further requiring plate-level excesses to be statistically significant, 13 of the 16 cells showed a significant excess on at least two different plates, and seven showed a significant excess on at least three plates (based on one-sided Poisson tests at an unadjusted $p < 0.05$). Thus, most observed transient excesses cannot be attributed to false detections on a single photographic plate.

Several of the 16 significant hotspots are spatially contiguous, forming broader geographic concentrations. The largest complex comprises nine interconnected cells, hotspots 1 - 3, 6, and 9 - 13, extending from approximately 105° to 125° W and from 5° to 25° N. This complex lies predominantly over the eastern Pacific, west of Mexico and Central America. A second complex comprises four adjacent cells, hotspots 4, 5, 7 and 8, spanning 90° - 100° W and 15° - 25° N. It encompasses parts of southern and eastern Mexico and the southern Gulf of Mexico, including the broader Tampico–Veracruz, Oaxaca, Tabasco–Chiapas and Bay of Campeche regions described above. The remaining three significant cells are spatially isolated from these two complexes: hotspot 14 in the northeastern Pacific at 137.5° W, 42.5° N; hotspot 15 in the southwestern United States (US) near Deming, White Sands and El Paso; and hotspot 16 farther south in the Pacific at 132.5° W, 12.5° N. Thus, when edge-sharing cells are grouped, the 16 significant hotspot cells form five spatially distinct regions: two multicell complexes and three isolated cells.

**Figure 1.**

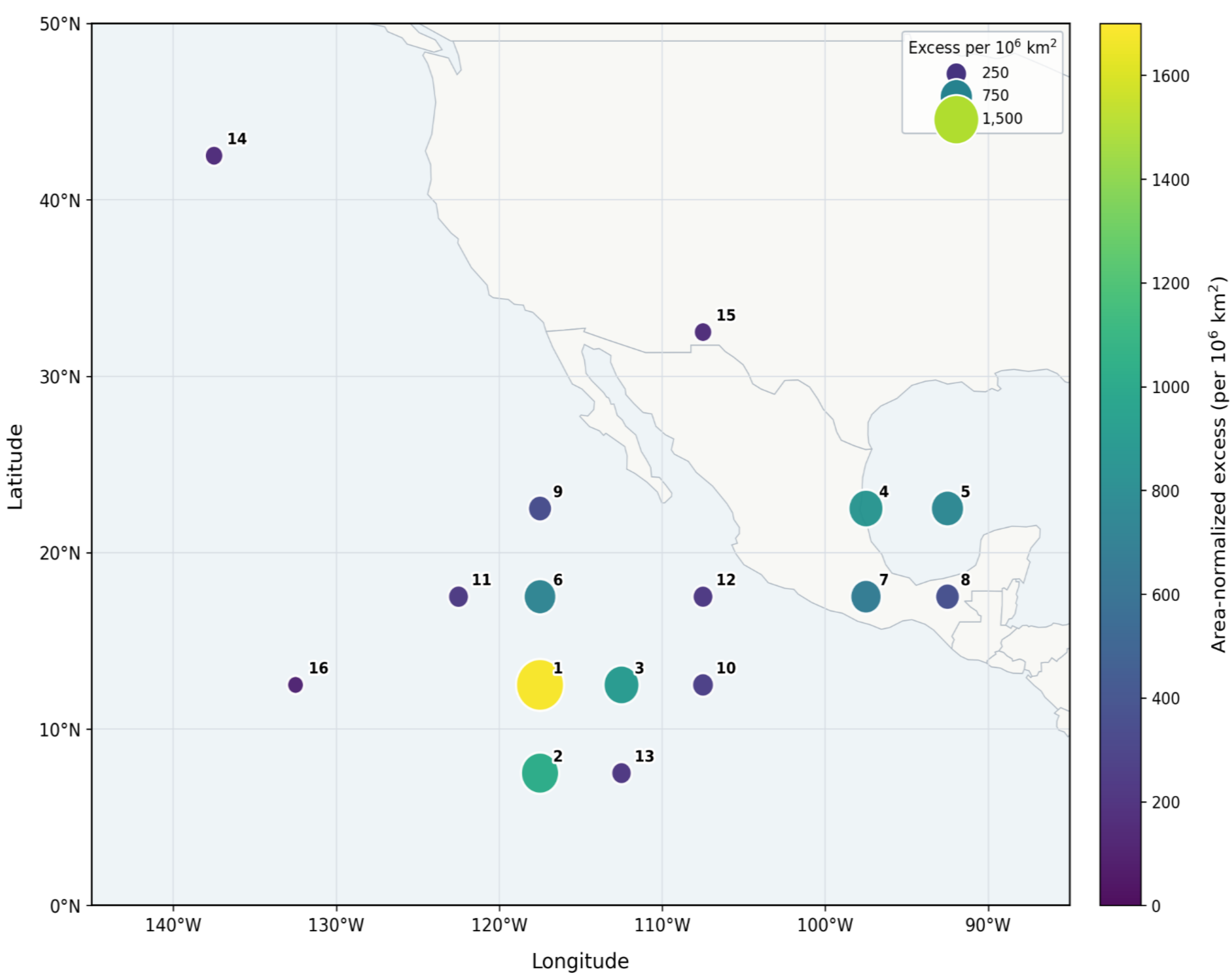

Cluster Analysis Results

*Transient Sample*

We next conducted two-step cluster analysis in a sample of n = 897 transients displaying an ML-assigned probability of ≥ 0.90 of being a real object rather than a plate defect (i.e., the highest confidence transients). This stringent criterion was chosen (rather than the ≥ 0.70 probability criterion used in the excess analysis) to minimize confounding influences of statistical noise (i.e., false positive transient identifications) on the clustering results. Cluster analysis in the transient sample produced 6 very distinct clusters, with locations highly similar within clusters and substantially different between clusters. The distinctiveness of the identified clusters is indicated by the observed average silhouette value of 0.70, which far exceeds the minimal criterion of 0.50 for indicating non-random clustering (see Methods). Relative percentages of the transient sample in each cluster and latitudes/longitudes of the centroids for each (i.e., average cluster location) are summarized in Table 2. Earth-projected locations of all 6 cluster centroids are displayed geographically in Figure 2 (labels for all cluster centroids are denoted with C as the first digit). Cluster 1 was centered in the Pacific Ocean to the southwest of Baja California. Clusters 5 and 6, while statistically distinct, were located in two adjacent regions of the Pacific further south than Cluster 1, with both being to the west of southern Mexico and Central America. These two clusters combined represented the location of nearly 65% of all high probability transients identified. Cluster 2, reflecting nearly a quarter of high probability transients, was centered in the Gulf of Mexico just off the coast of the Tampico-Vera Cruz area. Cluster 3 was located just southeast of Sedona, AZ, in the Sitgreaves National Forest. Cluster 4, reflecting the smallest proportion of the sample, was located in Jasper National Park, in British Columbia, Canada.

*Random Control Samples*

To test the possibility that the cluster analysis results above might simply be due to a random distribution or systematic observational biases in the POSS-I survey, we next conducted cluster analysis of three independent random location samples drawn from the excess analysis control dataset described above. Each was randomly selected to comprise n = 897 control locations, paralleling the size of the high probability transient sample. See Methods for details. If locations were truly random, cluster analyses should show silhouette values $<0.50$, a standard criterion for random clustering. If silhouette values $\geq 0.50$ were observed in these random control locations, any clustering would most likely be due to the POSS-I survey footprint, observing cadence, seasonal visibility, and sky coverage. If clusters reflecting similar locations were identified in both the real transient data and random control location data, this would suggest those specific transient clusters may represent artifacts of POSS-I observation parameters, unless supported by findings of the excess analysis that directly controlled for survey biases.

In the first random location dataset, three clusters were identified with an average silhouette value of 0.50, just meeting the minimal criterion for non-random clustering. Cluster 1 reflected 40.0% of the control sample and was centered on 18.1° N, 116.4° W. This location is southwest of Baja California in an area quite similar to Cluster 1 identified in the real transient data, suggesting the latter could potentially be an artifact of observing parameters (but see Discussion below). Cluster 2 in the first random location dataset comprised 37.6% of the sample and was in the Pacific Ocean off the coast of northern California (38.2° N, 124.8° W). There were no corresponding clusters in the real transient data. Finally, Cluster 3 in this random dataset represented 22.4% of this sample and was centered in the Northwest Territories of Canada (62.2° N, 121.4° W), again with no corresponding clusters in real transient data.

In the second random dataset, two clusters were identified with a silhouette value of 0.60, indicating stronger clustering than in the first random dataset but clustering weaker than in the high probability transient data (i.e., silhouette value = 0.70). Cluster 1 comprised 39.7% of the sample and was centered on 54.4° N, 120.8° W, in British Columbia, Canada, northwest of Jasper National Park. This area is in the same region as the smallest cluster (Cluster 4) identified in the real transient data, suggesting Cluster 4 in the transient data might be an artifact of observing biases. Cluster 2 in this second random location dataset reflected 60.3% of the random location sample and was centered on 23.5° N, 118.4° W. This region was due west of Baja California, an area similar to Cluster 1 in the first random dataset, and also broadly similar to Cluster 1 in the real transient data.

Finally, cluster analysis in the third random sample revealed two clusters with an average silhouette value of 0.60. Results were virtually identical to the second random sample, with one cluster located in British Columbia, Canada (54.1° N, 122.6° W) and the other located off the west coast of Baja California (22.4° N, 118.8° W).

<u>Is There Location Specificity of Transients Observed During Nuclear Weapons Tests?</u>

We next evaluated transient locations in the context of contemporaneous nuclear weapons testing. In prior work, we had found that transients were significantly increased within +/- 1 day of historic above ground nuclear tests [10,13]. We hypothesized that if these associations represented real rather than spurious associations, there would be evidence for location specificity of transients observed during nuclear weapons tests. That is, transients would be seen in the southwest US during nuclear tests conducted at the Nevada test site and would be seen over the Pacific during Pacific testing (i.e., Bikini Atoll, Enewetak).

To maximize likelihood of interpretable results, we again focused on the highest probability transients (≥ 0.90 probability of being a real transient per the ML model) [10]. There were 231 transients meeting this criterion observed within +/- 1 day of a nuclear test (for nuclear testing details, see https://nnss.gov/wp-content/uploads/2023/08/DOE_NV-209_Rev16.pdf). Of these, 203 transients (87.9%) were associated with a single nuclear test: Castle Yankee conducted on 5/4/54. This test evaluated a 13.5 megaton thermonuclear device (i.e., hydrogen bomb), the second largest ever tested by the US. It is notable that these 203 transients were all observed the night immediately before this test and were located exclusively over Pacific hotspot regions west of Mexico and Central America identified in hotspot analyses described above. None were observed over the southwestern U.S. or over any other regions identified in the excess analysis.

All other relevant nuclear weapons tests were conducted at the Nevada test site (5 tests of fission devices). Of the 28 transients associated with these tests, 24 were linked to a single test on 3/12/55, with one transient observed the night of the test (over Joshua Tree, CA immediately south of the test site) and 23 observed one night after the test (multiple locations over AZ and NM). Of the remaining 3 tests, associated transients were observed near

Kingman, AZ (1 transient), southwest of Baja California (2 transients), and Yukon, Canada (1 transient).

To test the location specificity hypothesis statistically, we created a 2 X 3 cross-tabulation matrix with two nuclear test locations (Pacific region, Southwestern U.S.) and three transient location regions (Pacific region, Southwestern U.S., and Other). Results of a Fisher's Exact Test ($p<.0001$) indicated significant clustering of the transients temporally associated with Pacific nuclear testing in the Pacific region, and clustering of the transients associated with testing at the Nevada test site located in the Southwest U.S. region (AZ, NM).

## Discussion

Using altitude estimates derived in our recent work [12], we estimated the Earth-projected locations of POSS-I transients. The aim of this study was to determine whether transient locations were randomly distributed, or if not, to identify specific hotspots above which transients appeared to group. There were both similarities and differences in results obtained between two independent statistical approaches.

Broadly, both methods found evidence for significant transient hotspots beyond what can be explained by random grouping or systematic biases related to POSS-I observation parameters. Using a method testing for excess transients relative to random expectations, 16 statistically-significant hotspots were identified. By comparison, cluster analysis identified 6 statistically-distinct hotspots. Similar cluster analyses in random control data suggested two of the cluster-derived hotspots might be artifacts of POSS-I observing parameters: one in British Columbia, Canada near Jasper and one in the Pacific southwest of Baja California. However, a similar hotspot southwest of Baja California was also identified in the excess analysis which directly controlled for POSS-I survey bias. Thus, while the Jasper hotspot remains questionable, results across the two methods suggest the hotspot southwest of Baja California may be real.

Detailed examination of locations identified across the two methods highlight interesting commonalities (see Table 3 and Figure 2). Nine of the 16 hotspots identified in the excess analysis, including the 3 hotspots with the largest statistical excess, were in the eastern Pacific due west of southern Mexico and Central America. This is the same region in which cluster analysis independently detected three statistical hotspots (Clusters 1, 5, and 6). On inspection of Figure 2 (combining results for both analyses), the dense grouping of transients using two independent methods in similar Pacific regions is quite striking, particularly given the near complete absence of hotspots identified by either method in the more northern Pacific region (hotspot 14 is an exception). Also particularly notable are findings that both methods identified hotspots in a similar region of the southwestern US (in the adjacent states of Arizona and New Mexico), with no hotspots identified anywhere else over the US mainland. Finally, both methods identified transient hotspots over the Gulf of Mexico between the coastal cities of Tampico and Vera Cruz. Excess analyses identified a broader complex of hotspots in this same region but further south and further east (i.e., southern Mexico and the Yucatan). There were no hotspots over land in more western or northern regions of Mexico.

**Figure 2.**

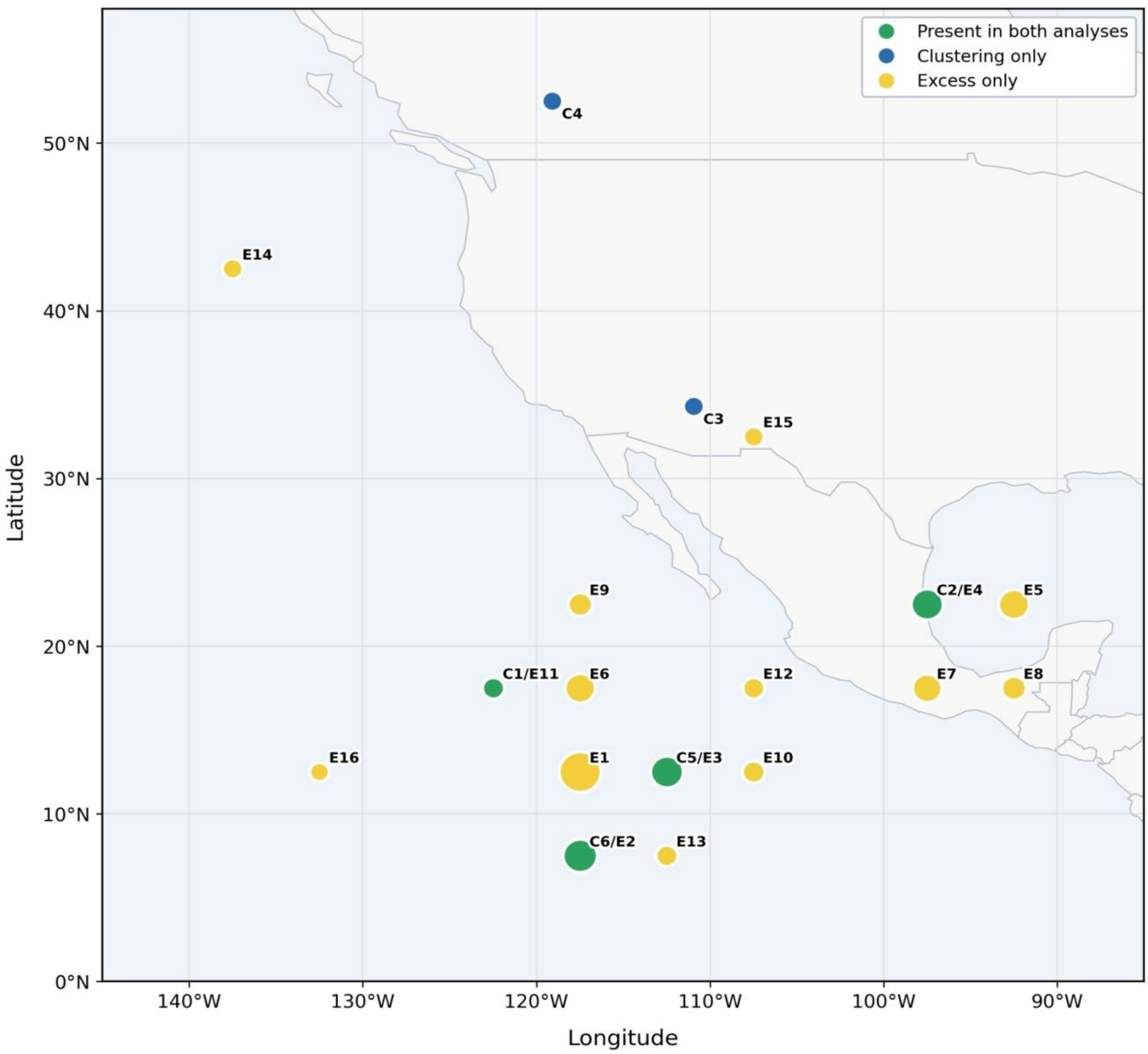

We propose that similar hotspots independently identified in both analyses are particularly informative for generating hypotheses regarding the nature of transients. In terms of prosaic explanations, for example, might the hotspots in the eastern Pacific be related in some way to seismic or unusual geomagnetic activity on the seafloor of that area? Notable geologic features associated with this general region include a portion of the East Pacific Rise and associated fault systems (Clipperton/Clarion) [16,17]. This region also has the highest known seafloor concentration of polymetallic nodules [18,19]. Hotspots identified in the southern Gulf of Mexico and the Yucatan were noted to be in the same general area as the Chicxulub asteroid impact, a region with documented geomagnetic and gravitational anomalies [20,22]. How such geological features might be linked to groupings of transients in Earth orbit is however unknown. We refer the reader to our prior work in which we have detailed and ruled out numerous other potential prosaic explanations for transients (e.g., cosmic rays, weather balloons, nuclear bomb debris, radiation contamination of photographic plates, Cherenkov-type phenomena) [10,13,14]. Our recent findings suggesting transients may be at approximately geosynchronous altitude[12] further constrain potential prosaic explanations.

While interpretative caution is warranted, the current work suggests that more provocative explanations may also warrant consideration. First, hotspots identified in nearby southwestern US regions using two independent statistical approaches are consistent with well-known narratives in the unidentified anomalous phenomenon (UAP) literature. The Sedona, AZ region (Cluster 3) is known as the top UAP sighting hotspot in the US [22,23]. This cluster is centered within Sitgreaves National Forest, the location of the famous Travis Walton UAP sighting and alleged abduction [24]. The nearby White Sands, NM region hotspot (identified via excess analysis) was the location of the first nuclear weapon test (the Trinity test); a transient hotspot in this area might be predicted given observed associations between transients and nuclear testing [10,13]. White Sands was also the location of frequent UAP sightings in the POSS-I era and was where early military rocket and guided missile tests were conducted [25-27]. We note however that rocket launches from this region cannot plausibly explain transients, given their observed point-like appearance (rather than streaks) during 50 minute POSS-I exposures. The Gulf of Mexico hotspot identified using both methods offshore from the Tampico-Vera Cruz region is an area with so many UAP sightings that many local residents believing there is an underwater "UFO base" offshore [28,29]. Finally, the hotspot closest to Baja California (E9, identified only using the excess method) is just south of the southern boundary of the US Navy W-291 training range where the famous "tic-tac" UAP incident occurred [30]. It is unknown whether the UAP-related location profiles above associated with these hotspots are meaningful or simply coincidental.

Beyond identifying transient location hotspots, we also evaluated specific Earth-projected locations for transients temporally associated with nuclear testing in prior work [10,13]. Results were consistent with this association displaying location specificity. Nearly 90% of the highest probability transients associated with nuclear testing (n = 203) were observed exclusively over Pacific region hotspots identified in this work, all *immediately prior to* the second largest ever US thermonuclear detonation, which was conducted further west in the Pacific. None of the transients associated with this test were located in identified hotspots over the US or Mexico. Conversely, no transients associated with Nevada nuclear tests were localized over the Pacific, and most (nearly 90%) were localized over the southwestern US, the hotspots in closest proximity to these tests.

Our prior published work supports high probability transients as real astronomical objects (rather than plate defects) that demonstrate specular reflections and were likely orbiting at approximately geosynchronous altitude prior to launch of the first artificial satellite [10,12,14]. The current study builds on this prior work and provides further evidence supporting the reality of high confidence transients, given their grouping in location hotspots in a non-random way, in some cases locations well-known in UAP lore. Extending previously reported temporal associations between transients and nuclear weapons tests [10,13], we found a correspondence between locations of nuclear tests and locations of associated transients. Finally, we further characterized our previous statistical findings that transients often appear shortly *before* nuclear tests [10]. While all prosaic explanations must be ruled out, one tentative interpretation of our findings could be that transients display characteristics suggesting possible intelligence, that is, a preference for specific locations over the Earth that may fit UAP lore and behavior suggesting possible awareness of imminent nuclear tests. While this hypothesis may fit available data, it is unfortunately not falsifiable. Replication of these findings in contemporaneous data from other observatories (e.g., Lick Observatory which is like Mount Palomar in California) would clarify the most appropriate interpretation of transient characteristics.

This study has potential limitations. Although unlikely, it is possible that observed transient hotspots were an artifact of POSS-I observational parameters, such as oversampling of specific locations. This possibility is mitigated by our examination of transient excesses relative to plate-aware controls with random locations that reproduce actual POSS-I plate coverage, permitting genuine transient excesses to be distinguished from apparent hotspots caused by uneven sky coverage. Another limitation is that a single plate containing an unusually large number of star-like plate defects (despite ML filtering) could produce an apparent hotspot in both the excess and cluster analyses. Because both use the same transient sample, their agreement alone does not rule out this possibility. However, secondary analyses found that most excess hotspots were supported by multiple independent plates, making it unlikely that defects on a single plate biased the results. A final limitation concerns the spatial extent and internal structure of identified hotspots. Because the excess analysis used fixed 5° × 5° cells, a broader geographic concentration of transients may be divided among several adjacent cells, while smaller or denser concentrations within a cell cannot be spatially resolved. The 16 significant hotspots should therefore not be interpreted as 16 independent or uniformly distributed physical hotspots. In particular, the two multicell complexes noted (e.g., in the eastern Pacific) may represent broad continuous regions, several smaller concentrations, or combinations of both. Likewise, the center of a significant cell should not be interpreted as the precise center of the underlying transient concentration. Replication using multiple cell sizes and shifted grid origins are needed to determine the true extent and localization of these hotspots.

In conclusion, POSS-I transients exhibit non-random clustering over distinct hotspots. Convergence in findings using two independent methods was observed for hotspots over the eastern Pacific off the west coast of southern Mexico and Central America, over the southwestern US (Sedona, AZ and White Sands, NM regions), and over the southern Gulf of Mexico. Replication of these findings with data from different observatories is warranted. Hotspot locations associated with nuclear weapons testing suggested a high degree of location specificity: localization over Pacific hotspots during Pacific nuclear testing and over southwestern US hotspots during testing at the Nevada test site. These findings further support the validity of the nuclear-transient association [10,13].

## Methods

### Transient Dataset

The initial source dataset was a sample of n = 107,875 transients derived by Solano et al. [1] and further detailed in Villarroel et al. [9]. For each analysis described below, we then selected from this full dataset only transients determined in our prior work [10] to have a probability of ≥0.70 (excess analysis) or ≥0.90 (cluster analysis) of being a real transient rather than a plate defect. These differing probability criteria were selected because binning the sample into 5°×5° cells (Earth projection) in the excess analysis required larger sample sizes, whereas optimal cluster analysis results would be obtained when statistical noise (e.g., due to false positives in the transient data) was minimized. We further selected only transients observed at declination > 3° and right ascension between 100° and 250°. These inclusion criteria were used to address empty stripes in the sky coverage caused by survey methodology and computational process [1,9].

This final filtered transient dataset contained not only right ascension and declination (J2000), but also time and date (Universal Time) for each transient observed. In a previous paper [12], altitude modelling based on quality-filtered transients in Table 1 suggested that transient altitudes potentially ranged from 20,000km to 35,000km. We note however that the largest shadow deficit occurred at 35,000km altitude, with Figure 2 revealing an excess of transients at low declinations [12]. Both of these latter findings are consistent with transients being at geosynchronous orbital altitude, an assumption we adopted for the current work. Based on the Palomar Observatory coordinates, we used line-of-sight projection geometry to project all transients onto a spherical shell at geosynchronous altitude using the celestial coordinates of each transient and the corresponding observation times. The resulting Earth-fixed locations are geocentric rather than geodetic, and all projected locations depended on the assumed fixed geosynchronous altitude. The same procedure was applied to the control sample, which represented 100 random points on each POSS-I plate covered by the VASCO transient survey [12,14]. By inspecting the control sample, we found that most projected positions are concentrated predominantly between 150° and 90° W in longitude, and 0° and 80° N in latitude. A small isolated group was also found near 140° E longitude and 45°-50° N in latitude. From the scatter map in the random control sample, we learned that spatial clustering often is caused by the survey footprint, observing cadence, seasonal visibility and sky coverage.

Calculation of Transient Excess

To control biases associated with the observational parameters above, we compared the observed number of transients with the number expected from the control sample derived from the same POSS-I plates included in the transient survey [12]. For the control sample, 100 random sky positions based on a uniform distribution were identified on each POSS-I plate, providing 63,500 control locations across the 635 POSS-I plates examined. Each control location was assigned to the observation date and time of its corresponding plate. Thus, control locations were yoked to the observation parameters of the POSS-I plates on which the transients were observed.

ChatGPT was used to assist with preparation of the transient and control samples for the excess analysis and the related Bonferroni-corrected conditional binomial tests. Both the transient and control samples were divided into nonoverlapping 5° × 5° location bins (cells) on Earth and we searched for an excess of transients relative to the control expectations in each cell. The final analysed sample had 11,418 transient locations and 23,814 random control locations. We assigned the projected positions for the transient and control datasets to identical 5° × 5° Earth-fixed longitude–latitude cells. For each cell, O denotes the observed number of transient candidates in the cell, C the number of control objects in that cell, and E the expected number of transients in that cell. The value C divided by 23,814 is the fraction of the control location sample falling in each cell, with C varying depending on the random distribution of control locations falling within each cell. Multiplying this fraction by the total number of transient locations gives the expected number of transients for that cell if they follow a random distribution. So, the number of expected transients for each cell is: $E = (C / 23{,}814) \times 11{,}418$. For example, if a cell contains 380 control locations, these represent about 1.60% of the control sample, so we expect approximately 1.60% of the 11,418 total transients to appear there by chance: $E = (380 / 23{,}814) \times 11{,}418 = 182.2$. If we actually observe 690 transients in that cell, this indicates $O/E = 3.79$. Whether there is a statistically significant transient excess was determined by a test of O versus E using conditional binomial tests. Note that using the methodology above, cells receiving more projected POSS-I survey coverage contain more control points and consequently have larger expected transient counts. This controls for any confounding of results due to POSS-I survey coverage patterns [9].

The random control sample occupied 152 cells (each 5° × 5°), of which 137 had at least 5 expected transients (statistical analyses were limited to these cells to provide a more conservative test of location clustering hypotheses). Among these 137 cells, 42 showed a positive excess ($O > E$), while 21 of these displayed significance of at least $2\sigma$ on the conditional binomial tests and were therefore classified as candidate hotspots. Because 137 cells were tested simultaneously, we then applied a conservative Bonferroni adjustment for multiple comparisons to all binomial test p values (i.e., nominal p value x 137 tests). These adjusted p values (Table 1) are directly interpretable relative to the chosen threshold of $p<.05$ for determining statistical significance.

To determine whether observed transient excesses were attributable to individual photographic plates, we decomposed each of the 16 significant 5° × 5° Earth-fixed cells according to the photographic plate on which each transient candidate and control object was detected. No plates or transient-rich observations were excluded. Importantly, the overlap between different plates is defined in the projected Earth-fixed coordinates and arises after projection onto the 35,786 km shell; it does not imply that the original photographic plates overlap in the sky.

For each plate contributing to a given cell, we calculated the observed number of high probability transient candidates, O, and the number of control objects, C. The plate-aware expected number of transients was calculated as described above. A plate was considered to contribute an excess to the cell when O > E. We then counted the number of distinct photographic plates independently contributing a positive excess to each hotspot (see Supplemental Table 1).

Cluster Analysis

Cluster analysis of Earth-projected transient locations was performed using the two-step cluster analysis procedure in IBM® SPSS® for Windows Version 31 (IBM; Armonk, New York). The two-step clustering approach is entirely data-driven and automatically selects the optimal number of clusters based solely on characteristics of the data [31,32]. This automated statistical cluster selection avoids potential bias in cluster selection related to prior expectations. In the current work, we employed the log-likelihood distance measure for determining clusters, and we used Schwarz's Bayesian Information Criterion (BIC) to determine the optimal number of clusters. The BIC is a standard criterion frequently used in statistics for optimizing model fit while also minimizing the risk of overfitting models (which can reduce the likelihood of the model replicating in an independent dataset).

The quality of a clustering model is indicated by the average silhouette value obtained [33,34]. The silhouette value is calculated for each case by comparing the average distance between an object and all other objects in the same cluster (a) with the average distance between that object and all objects in the nearest neighboring cluster (b). The silhouette value (s) is calculated as: $s = (b - a) / \max(a, b)$. Thus, the silhouette value is a measure of how similar an object is to its own cluster (cohesion) compared to other clusters (separation). Values can range from -1 to +1, with higher positive values indicating that the object is well matched to its own cluster and poorly matched to neighboring clusters. A cluster analysis producing an average silhouette value of 0.50 or greater indicates a good clustering model, reflecting clusters that are cohesive within themselves (i.e., objects within each cluster group together strongly based on locations) and that have good separation (i.e., the cluster locations are highly distinct from each other). Silhouette values < 0.50 indicate an inadequate clustering model, suggesting random data.

In preparation for cluster analysis, Earth-projected latitude and longitude (in degrees) of each transient meeting the criteria above (and random control locations) were converted to radians and then transformed into Cartesian (X,Y,Z) coordinates. This eliminates bias in the cluster analysis due to the spherical shape of the Earth, which results in distances (in degrees) between points near the poles being quite different than distances between points at the equator. To ensure accuracy, ChatGPT was employed to check the code used for these data transformations (AI was not used for any of the analyses themselves). Clustering variables in all analyses were the X, Y, Z Cartesian coordinates of each transient or control location (transient and control datasets were examined in separate analyses).

Using the highly stringent selection criteria noted above (ML probability ≥ 0.90, declination > 3°, right ascension between 100° and 250°), the final transient sample for cluster analysis comprised n = 897 very high probability transients with projected latitude and longitude estimates available at an assumed altitude of 35,786 km [12]. For comparability in cluster analyses, we randomly selected 3 independent samples each of n = 897 from the full random control location dataset used in the excess analysis. Cluster analysis generated centroids for each identified cluster, reflecting the average location for each cluster expressed as Cartesian coordinates. These centroids were saved, normalized to unit sphere, and converted back to latitude and longitude (in degrees) to permit identification of the corresponding geographic location using Google maps.

Clustering results for the real transient dataset were interpreted in context of the cluster analysis findings for the three random control samples. That is, similar clusters identified in both the real transient data and the random control data were interpreted as potentially related to the effects of spatial clustering due to the POSS-I survey footprint, observing cadence, seasonal visibility, and sky coverage, rather than representing a grouping of real transients.

## Acknowledgments

The authors would like to express their appreciation to Dr. Enrique Solano for his work creating the original transient identification pipeline that produced the full transient dataset from which the machine learning data examined in the current work was derived. The authors would also like to express their deep gratitude to N. Colosimo for the extensive technical input, codes, calculations, and conceptual insights instrumental for estimating POSS-I transient altitudes that enabled derivation of Earth-projected transient latitudes and longitudes that were the focus of the current work.

Author Contributions

S.B. helped design the study, assisted in conducting and interpreting the statistical analyses, and prepared a draft of the initial manuscript.

B.V. helped design the study, assisted in creating the original transient dataset, assisted in conducting and interpreting the statistical analyses, and assisted in preparing a draft of the initial manuscript.

H.G. conducted the independent verification of the Earth-shadow analysis, developed and validated the line-of-sight projection calculations, and assisted in editing the final version of the manuscript.

B.D. compiled the datasets, created the machine learning model used to identify high probability transients, and assisted in editing the final version of the manuscript.

A.S. assisted in conducting and interpreting the statistical analyses and editing the final version of the manuscript.

Funding

B.V. is funded by the Swedish Research Council (Vetenskapsr\aa det, grant no. 2024-04708) and supported by a generous donor. A.S. is supported by the Athanatos Foundation.

Data Availability Statement

The final dataset reflecting all analyzed variables will be made available by the authors upon reasonable request to Dr. Beatriz Villarroel (beatriz.villarroel@su.se).

Competing Interest Statement

None of the authors have competing interests to declare.

Figure Legends

Figure 1. Geographical locations of the 16 hotspots as identified by significant transient excess (labeled in descending order of the magnitude of the transient excess). Larger sized hotspots indicate greater excesses of transients compared to expectations based on random location controls.

Figure 2. Comparison of the clustering and excess analyses. Hotspots identified in the excess analysis only are in portrayed in yellow, hotspots identified in the cluster analysis only are displayed in blue, and hotspots identified in both analyses within the same 5° × 5° location cell are displayed in green.

Table 1. Hotspot cells at an assumed altitude of 35,786 km displaying a significant excess of transeints relative to random expectations based on bonferroni-corrected conditional binomial tests. The cells are ordered by area-corrected excess density. Here, O and E denote the observed and expected transient counts, respectively; Zbin is the Gaussian-equivalent Z score corresponding to the one-sided conditional-binomial test for transient excess; pBonf is the Bonferroni-corrected p-value for the conditional-binomial test. pBonf values < .05 are considered significant after controlling for familywise error.

| **Hotspot Number** | **Hotspot Cell center** | **O / E** | **Ratio** | **Excess per $10^6$ km²** | **Zbin** | **pBonf** |
|---|---|---|---|---|---|---|
| 1 | 117.5° W, 12.5° N | 690 / 182.2 | 3.79 | 1,683.2 | 21.43 | < .0001 |
| 2 | 117.5° W, 7.5° N | 414 / 98.8 | 4.19 | 1,028.9 | 17.45 | < .0001 |
| 3 | 112.5° W, 12.5° N | 455 / 182.7 | 2.49 | 902.7 | 13.06 | < .0001 |
| 4 | 97.5° W, 22.5° N | 258 / 14.4 | 17.94 | 853.3 | 20.24 | < .0001 |
| 5 | 92.5° W, 22.5° N | 231 / 13.9 | 16.61 | 760.4 | 18.94 | < .0001 |
| 6 | 117.5° W, 17.5° N | 420 / 204.3 | 2.06 | 732.1 | 10.33 | < .0001 |
| 7 | 97.5° W, 17.5° N | 217 / 21.1 | 10.29 | 664.7 | 16.87 | < .0001 |
| 8 | 92.5° W, 17.5° N | 123 / 12.5 | 9.87 | 375.1 | 12.55 | < .0001 |
| 9 | 117.5° W, 22.5° N | 293 / 192.3 | 1.52 | 352.8 | 5.37 | < .0001 |
| 10 | 107.5° W, 12.5° N | 219 / 136.2 | 1.61 | 274.6 | 5.17 | < .0001 |
| 11 | 122.5° W, 17.5° N | 278 / 206.2 | 1.35 | 243.7 | 3.80 | 0.0098 |
| 12 | 107.5° W, 17.5° N | 228 / 157.7 | 1.45 | 238.4 | 4.18 | 0.002 |
| 13 | 112.5° W, 7.5° N | 141 / 70.5 | 2.00 | 230.2 | 5.76 | < .0001 |
| 14 | 137.5° W, 42.5° N | 53 / 13.9 | 3.81 | 171.6 | 5.86 | < .0001 |
| 15 | 107.5° W, 32.5° N | 133 / 89.2 | 1.49 | 168.1 | 3.42 | 0.0431 |
| 16 | 132.5° W, 12.5° N | 74 / 36.9 | 2.00 | 122.9 | 4.15 | 0.0023 |

Table 2. Results of two-step cluster analysis of locations of n = 897 POSS-I transients with a machine-learning derived probability of ≥ 0.90 of being a real object. Cluster centroids are the average location of objects in each cluster.

| **Cluster Number** | **Percentage of Sample** | **Cluster Centroid Longitude/Latitude** |
|---|---|---|
| 1 | 9.4 | 122.3° W, 17.0° N |
| 2 | 21.5 | 95.5° W, 20.3° N |
| 3 | 3.2 | 110.9° W, 34.3° N |
| 4 | 2.6 | 119.1° W, 52.5° N |
| 5 | 25.8 | 111.5° W, 14.1° N |
| 6 | 37.6 | 117.1° W, 10.0° N |

Table 3. Comparison between the six transient-cluster centroids and their nearest significant excess cells. Four cluster centroids fall within the same 5° × 5° cell as an excess hotspot.

| Transient Cluster | Cluster Centroid Longitude/Latitude | Nearest Excess Hotspot | Hotspot Cell Center | Separation (km) | Same 5° × 5° Cell? |
|---|---|---|---|---|---|
| 1 | 122.3° W, 17.0° N | 11 | 122.5° W, 17.5° N | 57 | Yes |
| 2 | 95.5° W, 20.3° N | 4 | 97.5° W, 22.5° N | 319 | Yes |
| 3 | 110.9° W, 34.3° N | 15 | 107.5° W, 32.5° N | 377 | No |
| 4 | 119.1° W, 52.5° N | 14 | 137.5° W, 42.5° N | 1,766 | No |
| 5 | 111.5° W, 14.1° N | 3 | 112.5° W, 12.5° N | 211 | Yes |
| 6 | 117.1° W, 10.0° N | 2 | 117.5° W, 7.5° N | 281 | Yes |

Supplemental Table 1. Number of distinct POSS-I photographic plates contributing projected transient candidates to each significant Earth-fixed 5° × 5° cell, the number of plates individually showing a positive projected excess (O > E), and the number showing a statistically significant excess (one-sided Poisson unadjusted $p < 0.05$). Importantly, overlap between different plates is defined in the projected Earth-fixed coordinates after projection onto the 35,786 km shell and does not imply overlap between the original photographic footprints in the sky. Fourteen of the 16 cells show a positive projected excess originating from more than one photographic plate.

| **Hotspot Cell Center** | **O / E** | **Ratio** | **Number of Plates Contributing Transients** | **Number of Plates with O > E** | **Number of Plates with Significant O > E** |
|---|---|---|---|---|---|
| 117.5° W, 12.5° N | 690 / 182.2 | 3.79 | 15 | 11 | 9 |
| 117.5° W, 7.5° N | 414 / 98.8 | 4.19 | 7 | 3 | 3 |
| 112.5° W, 12.5° N | 455 / 182.7 | 2.49 | 14 | 7 | 6 |
| 97.5° W, 22.5° N | 258 / 14.4 | 17.94 | 3 | 3 | 2 |
| 92.5° W, 22.5° N | 231 / 13.9 | 16.61 | 1 | 1 | 1 |
| 117.5° W, 17.5° N | 420 / 204.3 | 2.06 | 17 | 12 | 9 |
| 97.5° W, 17.5° N | 217 / 21.1 | 10.29 | 3 | 3 | 2 |
| 92.5° W, 17.5° N | 123 / 12.5 | 9.87 | 1 | 1 | 1 |
| 117.5° W, 22.5° N | 293 / 192.3 | 1.52 | 16 | 7 | 6 |
| 107.5° W, 12.5° N | 219 / 136.2 | 1.61 | 9 | 4 | 2 |
| 122.5° W, 17.5° N | 278 / 206.2 | 1.35 | 13 | 6 | 4 |
| 107.5° W, 17.5° N | 228 / 157.7 | 1.45 | 11 | 3 | 2 |
| 112.5° W, 7.5° N | 141 / 70.5 | 2.00 | 4 | 3 | 3 |
| 137.5° W, 42.5° N | 53 / 13.9 | 3.81 | 2 | 2 | 1 |
| 107.5° W, 32.5° N | 133 / 89.2 | 1.49 | 8 | 3 | 2 |
| 132.5° W, 12.5° N | 74 / 36.9 | 2.00 | 3 | 2 | 2 |